\documentclass[final,english]{bullsrsl}

\usepackage[normalem]{ulem}
\usepackage[latin1]{inputenc}
\usepackage[T1]{fontenc}
\usepackage{subcaption}
\usepackage{aadefs} 
\usepackage{natbib} 
\usepackage{graphicx}
\usepackage{amsmath}	
\usepackage{amssymb}

\begin{document}
\title{Optical spectroscopy of comets}

\author[corresponding]{K}{Aravind}
\author{Shashikiran}{Ganesh}
\affiliation{Physical Research Laboratory, Ahmedabad 380009, India}
\correspondance{aravind139@gmail.com}

\maketitle


%

\begin{abstract}
Comets are pristine remnants of the Solar system, composed of dust and ice. They remain inactive and undetectable for most of their orbit due to low temperatures. However, as they approach the Sun, volatile materials sublimate, expelling dust and creating a visible coma. Spectroscopic observations of comets help the simultaneous study of both the gas emissions and reflected sunlight from dust particles. By implementing a long slit, the spatial variations in molecular emissions can be analysed to be further used for other computations. Additionally, spatial information aids in extracting the characteristic profile of the Af$\rho$ parameter, revealing insights into the behaviour of dust emissions. A sufficiently long slit would prove advantageous in extracting information about the emissions occurring at different parts of the coma or even the tail. We can gain an overall comprehensive understanding of a comet's chemical composition and dust emission by constructively utilising low-resolution spectroscopy with the help of a long slit.
\end{abstract}

\keywords{Solar system minor bodies, Comet, Optical spectroscopy}



\section{Introduction}
Comets, heavenly objects of awe, have been captivating observers for centuries with their stunning displays in the night sky. These celestial wanderers, composed of icy material and dust particles, traverse vast interplanetary distances. Scientifically intriguing and visually enchanting, comets provide valuable insights into the formation and evolution of our Solar System.\\
Comets are remnants of the early stages of the Solar System, carrying within them the pristine material that dates back to its birth approximately 4.6 billion years ago. These icy bodies primarily originate from two distinct regions: the Kuiper Belt, a disk-shaped region beyond the orbit of Neptune \citep{reservoir_formation}, and the Oort Cloud, a hypothetical enormous spherical shell located at the outermost edges of the Solar System \citep{jan_oort}.\\
The nucleus, the solid core of a comet, acts as the reservoir of volatile substances such as water, carbon dioxide, methane, and ammonia \citep{Making_comet_nucleus}, encapsulated within a matrix of rock, dust, and organic compounds \citep{organic_volatile_compo_mumma}. As these bodies spend the majority of their time in orbit away from the Sun, they would have undergone minimal internal and external evolution. Hence, studying comets provides us with invaluable insights into the early conditions of the Solar System. Cometary compositions reflect the chemical abundances present during the formation of the Sun and the planets, offering a time capsule of pristine material that has remained relatively unaltered over billions of years.\\
As a comet approaches the Sun on its elliptical orbit, heat and radiation cause these volatile substances to sublimate, generating a fuzzy coma-a gaseous and dusty envelope-around the nucleus \citep{swamy}. Additionally, the solar wind and radiation pressure sweep away some escaping particles, forming the iconic tail that points away from the Sun. By analysing the relative abundances of elements within comets, we can decipher the similarities/dissimilarities of composition between comets originating from different reservoirs, shedding light on the processes that led to the diversity and complexity of planetary systems.\\
In recent years, space missions such as Rosetta \citep{rosetta} and Stardust \citep{stardust} have ventured close to comets, providing unprecedented opportunities for detailed investigations. These missions have collected invaluable data about comet nuclei, their surface structures, compositions, and gas dynamics, significantly enhancing our understanding of these enigmatic objects. However, it is not practical to send space missions to all possible comets. Hence, ground-based observations utilising various techniques play a major role in having continuous monitoring of these icy bodies originating from different reservoirs to improve our understanding further. \\
This article aims to present a comprehensive overview of 
the benefits of using optical spectroscopic methods in exploring the compositional characteristics of comets. Along with this, we aim to put forward the advantage of long-slit spectroscopy in unlocking the mysteries of these captivating celestial wanderers and advancing our understanding of the cosmos.

\section{Observation and Reduction}
\begin{figure}
\centering
\includegraphics[scale=0.6]{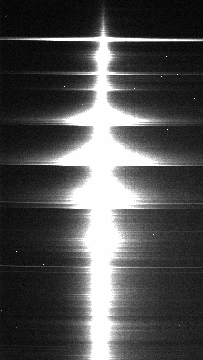}
  \bigskip
  \hspace{1cm}
  \includegraphics[scale=0.4]{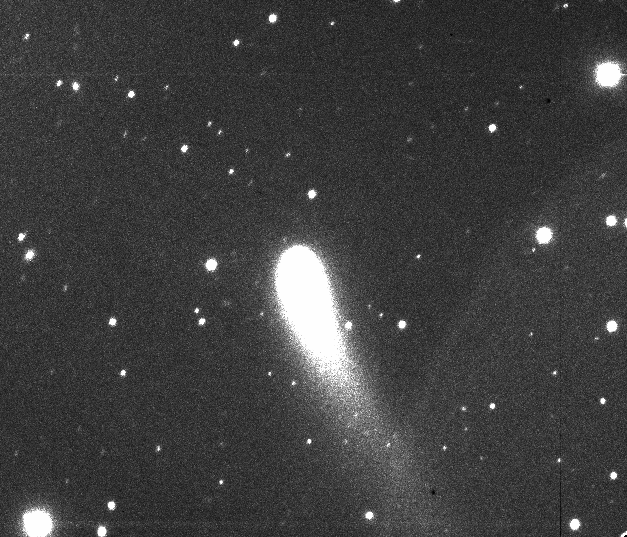}
\begin{minipage}{12cm}
\caption{\textit{Left panel} illustrates the raw data of the spectroscopic observation of comet C/2020 F3 (NEOWISE) 
 and the \textit{Right panel} illustrates the raw data of imaging observation of comet C/2022 E3 (ZTF), observed from HCT, Hanle.}
\label{rawdata}
\end{minipage}
\end{figure}

Cometary observations are majorly carried out from the three main observatories in India, namely, the 1.2~m telescope at Mount Abu InfraRed Observatory (MIRO) operated by the Physical Research Laboratory at Mount Abu, Rajasthan (Longitude: 72$^\circ$~46$^\prime$~47.5$^{\prime \prime}$ East;  Latitude: 24$^\circ$~39$^\prime$~8.8$^{\prime \prime}$ North; Altitude: $1680~$m), the 2~m Himalayan Chandra Telescope (HCT) operated by the Indian Institute of Astrophysics at Hanle, Ladakh (Longitude: 78$^\circ$~57$^\prime$~49.8$^{\prime \prime}$ East; Latitude: 32$^\circ$~46$^\prime$~46.3$^{\prime \prime}$ Altitude: $4475$~m) and the 3.6 m Devasthal Optical Telescope (DOT) situated at Devasthal operated by the Aryabhatta Research Institute of Observational Sciences, Nainital (Longitude: 79$^\circ$~41$^\prime$~04$^{\prime \prime}$ East;  Latitude: 29$^\circ$~31$^\prime$~39$^{\prime \prime}$ North; Altitude: $2540~$m). While these telescopes are primarily used for Spectroscopic observations and Imaging, Polarimetric observations are also carried out frequently as additional observations. The raw data of cometary observation is illustrated in Figure \ref{rawdata}. Self-scripted \textsc{Python} codes and different modules in \textsc{IRAF} are employed to perform the preliminary reductions and calibration (both wavelength and flux) to get the data ready for further scientific analysis.

\section{Spectroscopic analysis of cometary compositions.}
\begin{figure}[h!]
    \centering
    \includegraphics[width=0.8\linewidth]{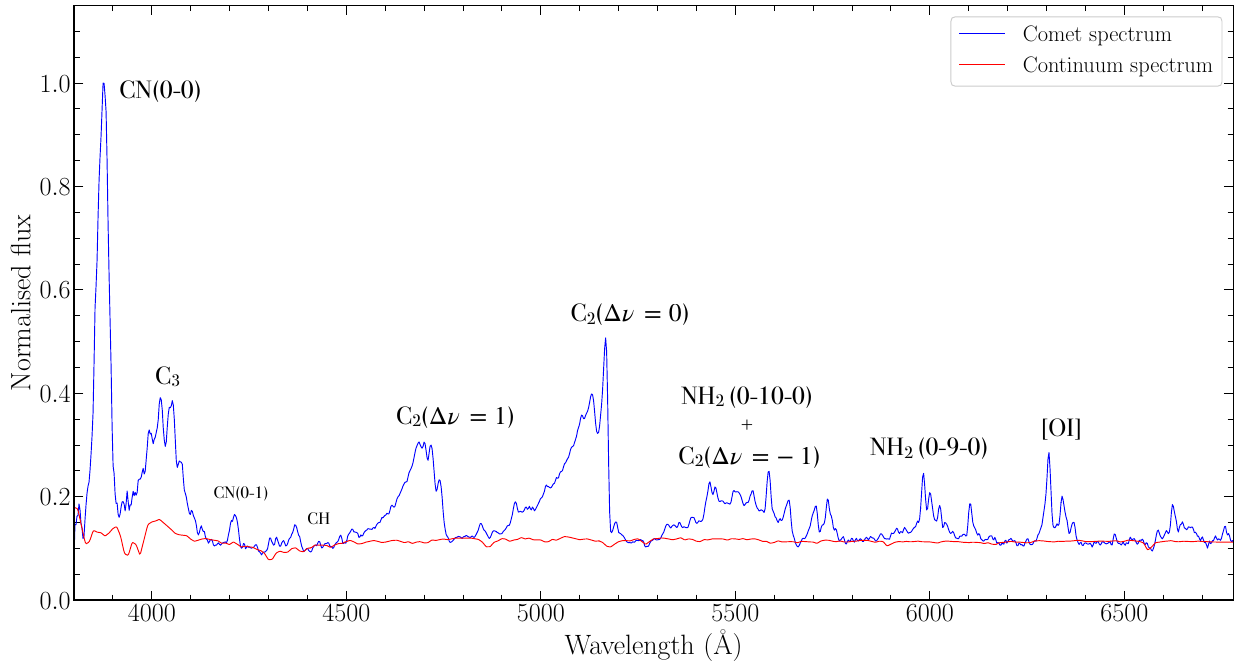}
    \begin{minipage}{12cm}
    \caption{Optical spectrum of comet 46P/Wirtanen observed from Mt. Abu observatory.}
    \label{fig:46P}
    \end{minipage}
\end{figure}

\begin{figure}
\centering
\includegraphics[scale=0.43]{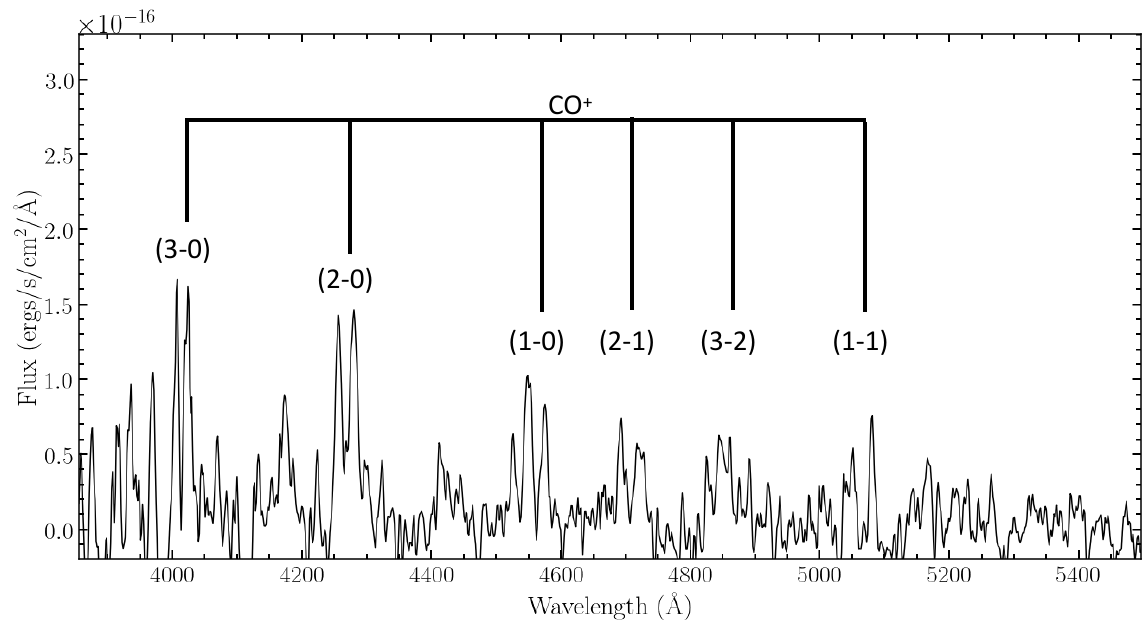}
  \bigskip
  \includegraphics[scale=0.24]{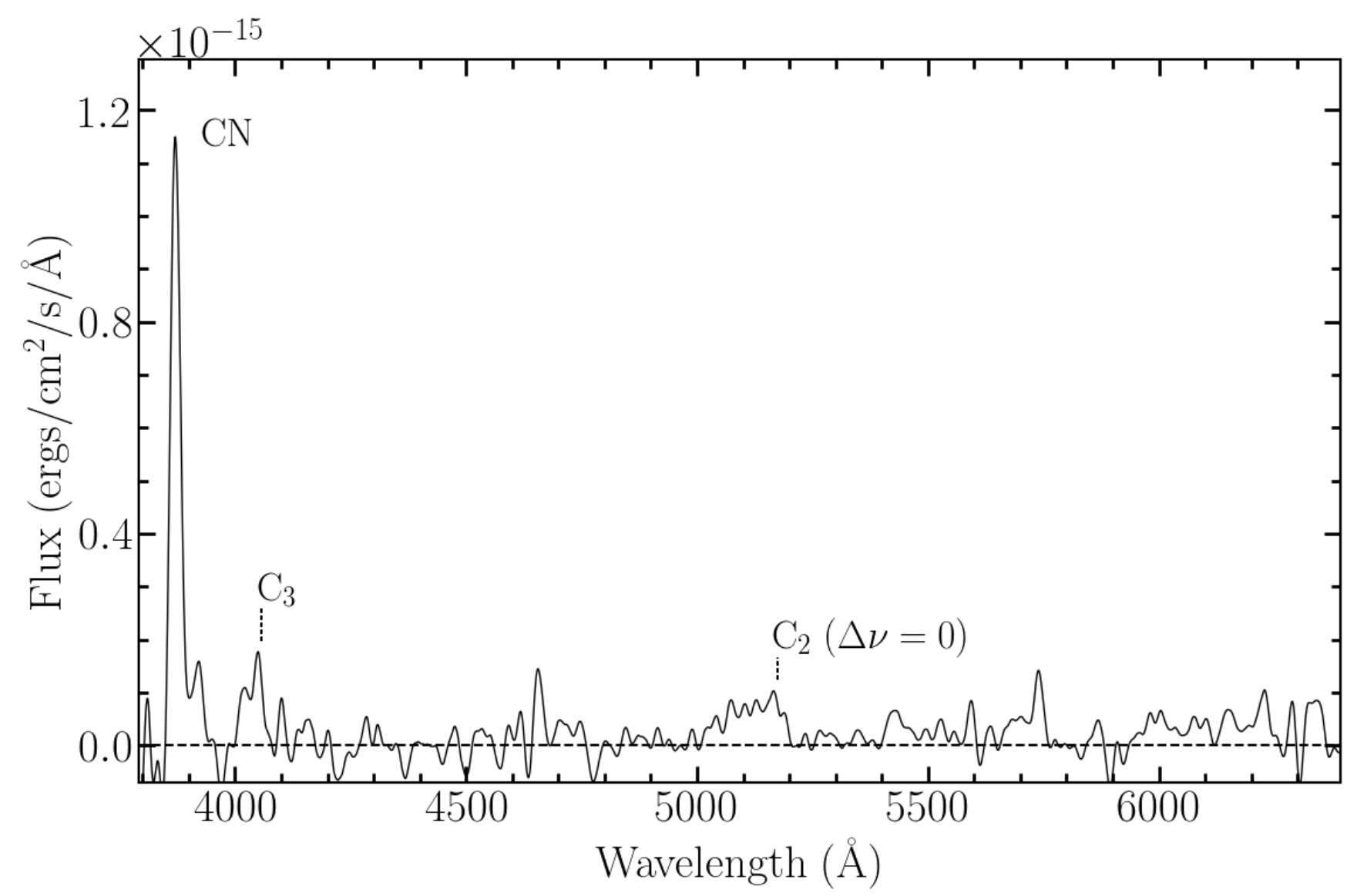}
\begin{minipage}{12cm}
\caption{\textit{Left panel} illustrates the optical spectrum of the comet 29P/Schwassmann-Wachmann and the \textit{Right panel} illustrates the optical spectrum of interstellar comet 2I/Borisov.}
\label{spectrum_comp}
\end{minipage}
\end{figure}

When it comes to unveiling the morphological or compositional properties of a comet, there exist multiple observational techniques, namely Imaging (both Broad band and Narrow band), Polarimetry (Broad band) and Spectroscopy (Low resolution and High resolution). While Polarimetry is particularly used for understanding the physio-compositional properties of the dust present in the comets, both Imaging and Spectroscopy can be effectively employed to extract information concerning both dust and gaseous material.\\
Spectrum is simply the variation of intensity with wavelength. Spectroscopic observations provide us with an opportunity to probe the various molecular emissions present in the cometary spectrum. A comet spectrum is comprised of both the coma gas emission spectrum (fluorescence emission) and the spectrum of the Sunlight scattered by the dust particles (continuum). The optical regime of the cometary spectrum is filled with emissions from different radicals, which are the daughter molecules of the parent ice species present in the comet nucleus. Figure \ref{fig:46P} illustrates a typical spectrum of a comet with clear detection of emissions from different bands of various radicals like $CN (\Delta \nu = 0$; band head at $\lambda3880$\AA ), $C_3 (\lambda4050$\AA), $C_2 (\Delta \nu = 0$; band head at $\lambda5165$\AA ) and NH$_2$ (multiple bands above $\sim$ 5000 \AA) etc., along with the forbidden atomic Oxygen lines, [OI] (5577 \AA, 6300 \AA ~and 6364 \AA). Spectroscopic analysis of this wavelength region along the orbit of a comet can provide immense details regarding the relative abundance of different molecular emissions and their variations with heliocentric distances.\\
\begin{figure}
\centering
\includegraphics[scale=0.22]{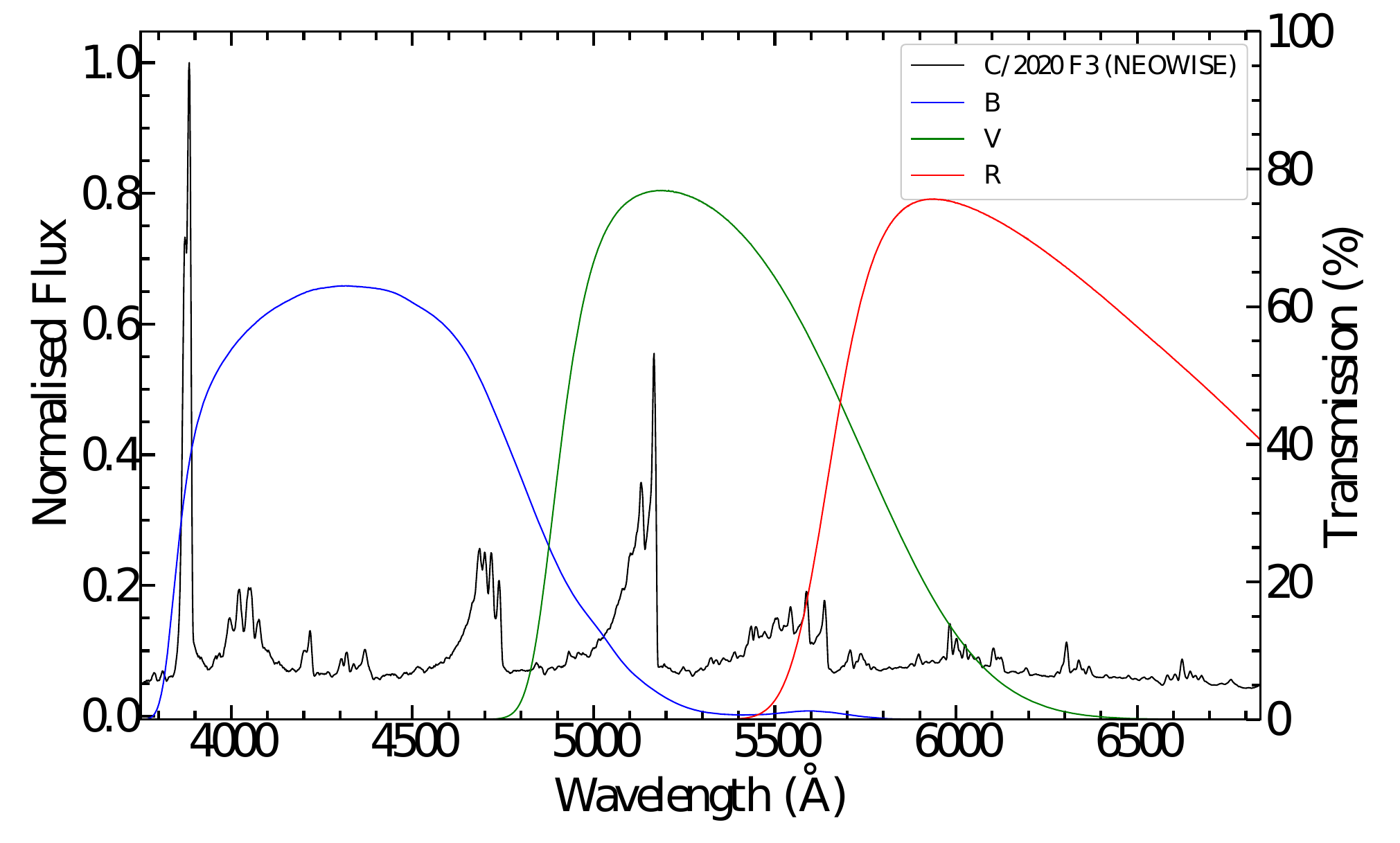}
  \bigskip
  \includegraphics[scale=0.22]{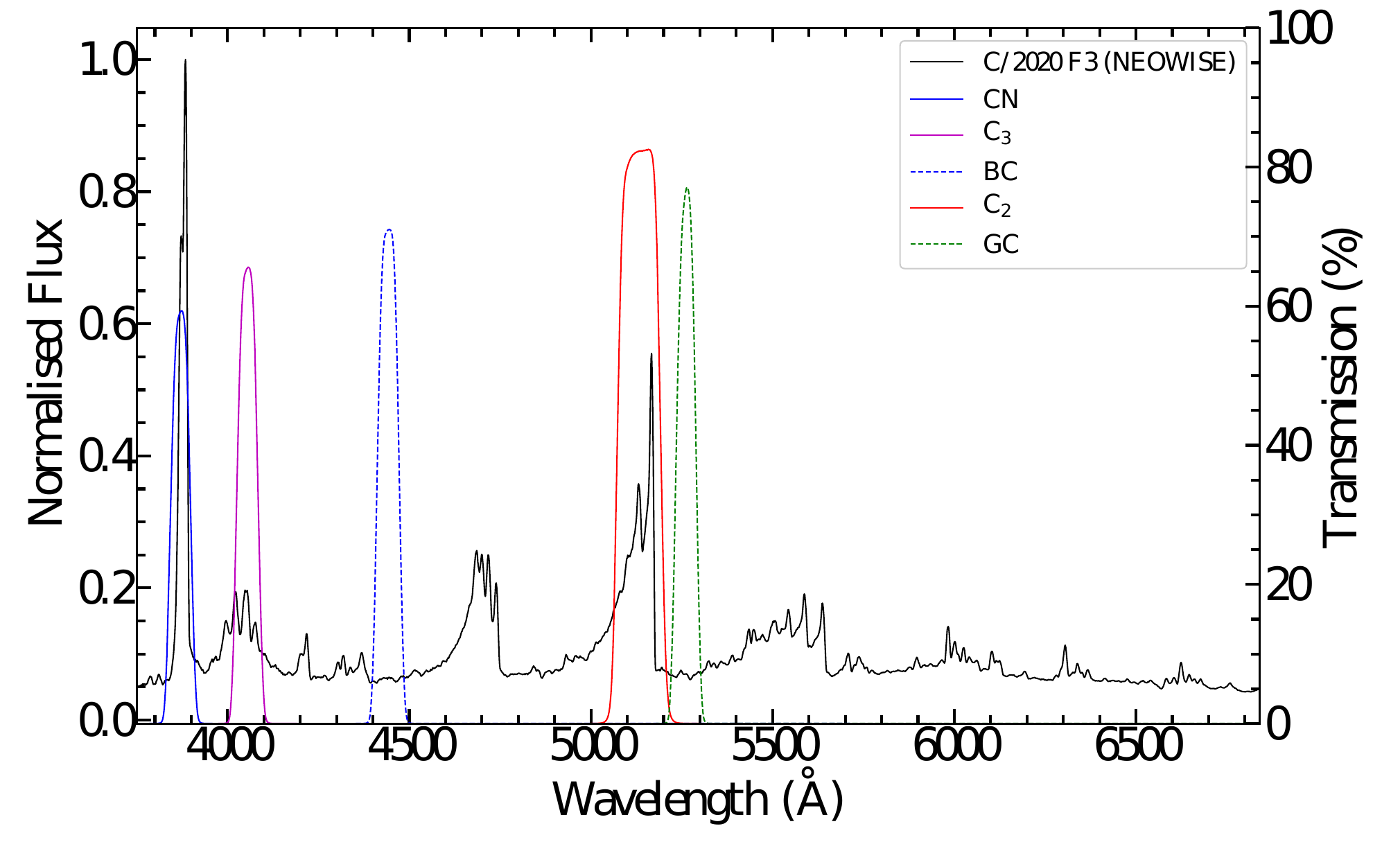}
\begin{minipage}{12cm}
\caption{The optical spectrum of comet C/2020 F3 observed from HCT overplotted with the transmission profiles of a few Broad band filters (\textit{left panel}) and of a few Narrow band filters (\textit{right panel}).}
\label{comet_imaging}
\end{minipage}
\end{figure}
Imaging observations of a comet consists of two methods, one using the Broad band filters as in \textit{UBVRI} \citep{bessell}, and the other using Narrow band filters as those defined in \cite{HB}. As shown in the left panel of Figure \ref{comet_imaging}, the bandwidth of any broad band filter would contain contamination from two or more molecular emission bands, making it effective in only analysing the morphological features present in the coma or the rough  nucleus rotation from the light curve. The \textit{R} or \textit{I} bands, least contaminated by the molecular emissions, can be used to analyse the dust production through a proxy parameter known as Af$\rho$ (product of albedo (A), filling factor of the grains within our field of view (f), and the linear radius of the field of view at the
comet ($\rho$)) as defined in \cite{Ahearn_Bowel_slope}. \\
In the mid-1990s, when the comet Hale-Bopp was discovered, a set of Narrow band filters (OH, NH, CN, C$_3$, C$_2$, CO$^+$, UV continuum, Blue continuum, Green continuum and Red continuum) were designed, hence known as the Hale-Bopp filters, to isolate the major molecular and dust emissions  present in comets (see right panel of Figure \ref{comet_imaging} for the transmission profiles of a few Narrow band filters). Even though the Narrow band filters facilitate fast observations and better Signal-to-Noise Ratio (SNR) in the case of faint comets, the advantage of spectroscopic analysis is the luxury of obtaining all the emissions present in a given wavelength range in a single shot avoiding the effects of any temporal variation, if present.\\
Spectroscopy is also advantageous in visualising the similarity/dissimilarity of emissions occurring in a comet in comparison to other observed comets. Comets 29P/Schwassmann-Wachmann (see left panel of Figure \ref{spectrum_comp}) and C/2016 R2 \citep{Kumar_R2} are two such comets where spectroscopy was effectively used to visualise the diversity in the emissions arising from them. Similarly, it was possible with spectroscopy to establish that the first interstellar comet 2I/Borisov had molecular emissions similar to those observed in comets of our Solar system \citep{borisov_aravind} (see right panel of Figure \ref{spectrum_comp}).

\subsection{Benefits of long slit spectroscopy}
\begin{figure}
\centering
\includegraphics[scale=0.245]{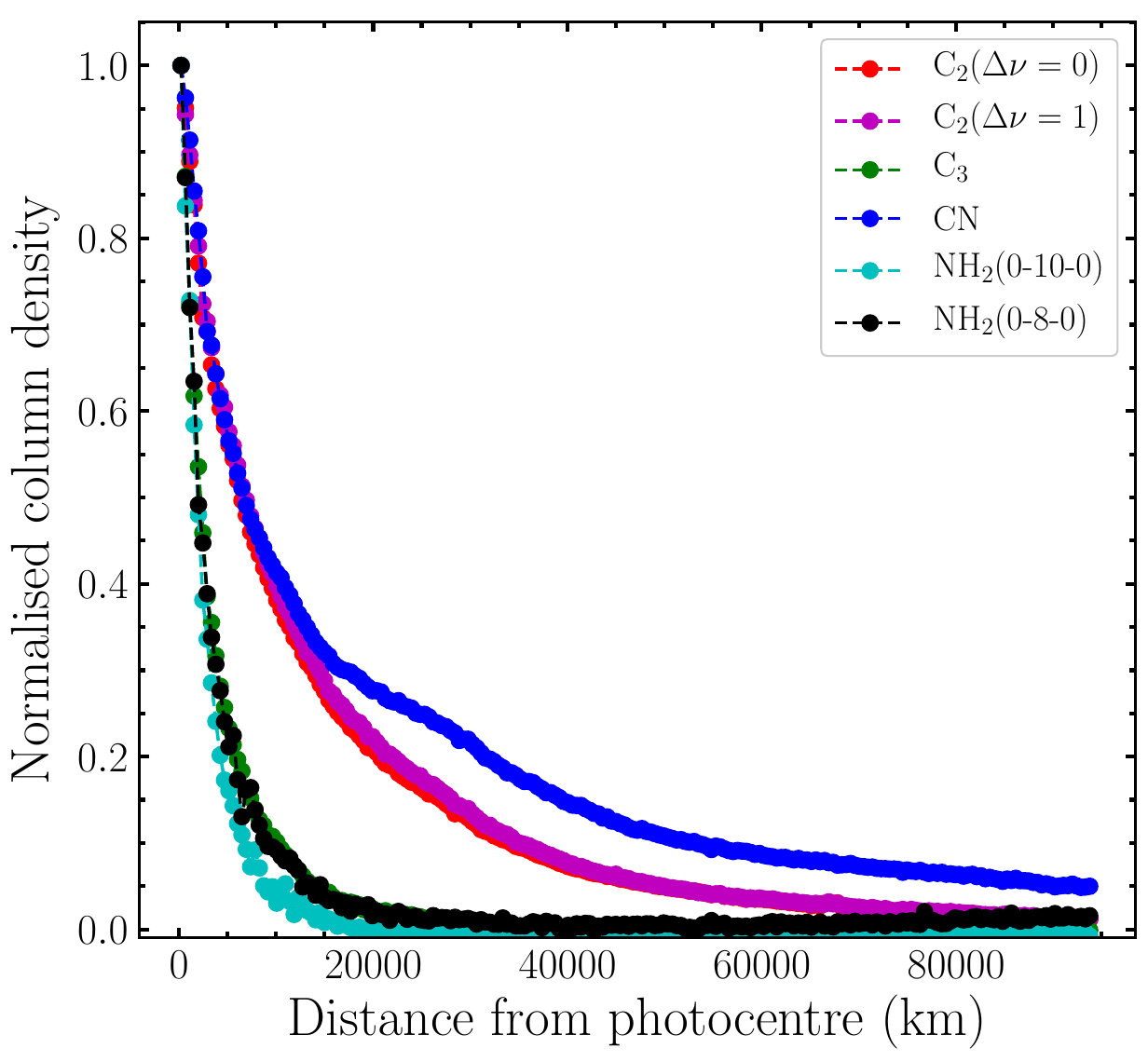}
  \bigskip
  \includegraphics[scale=0.2]{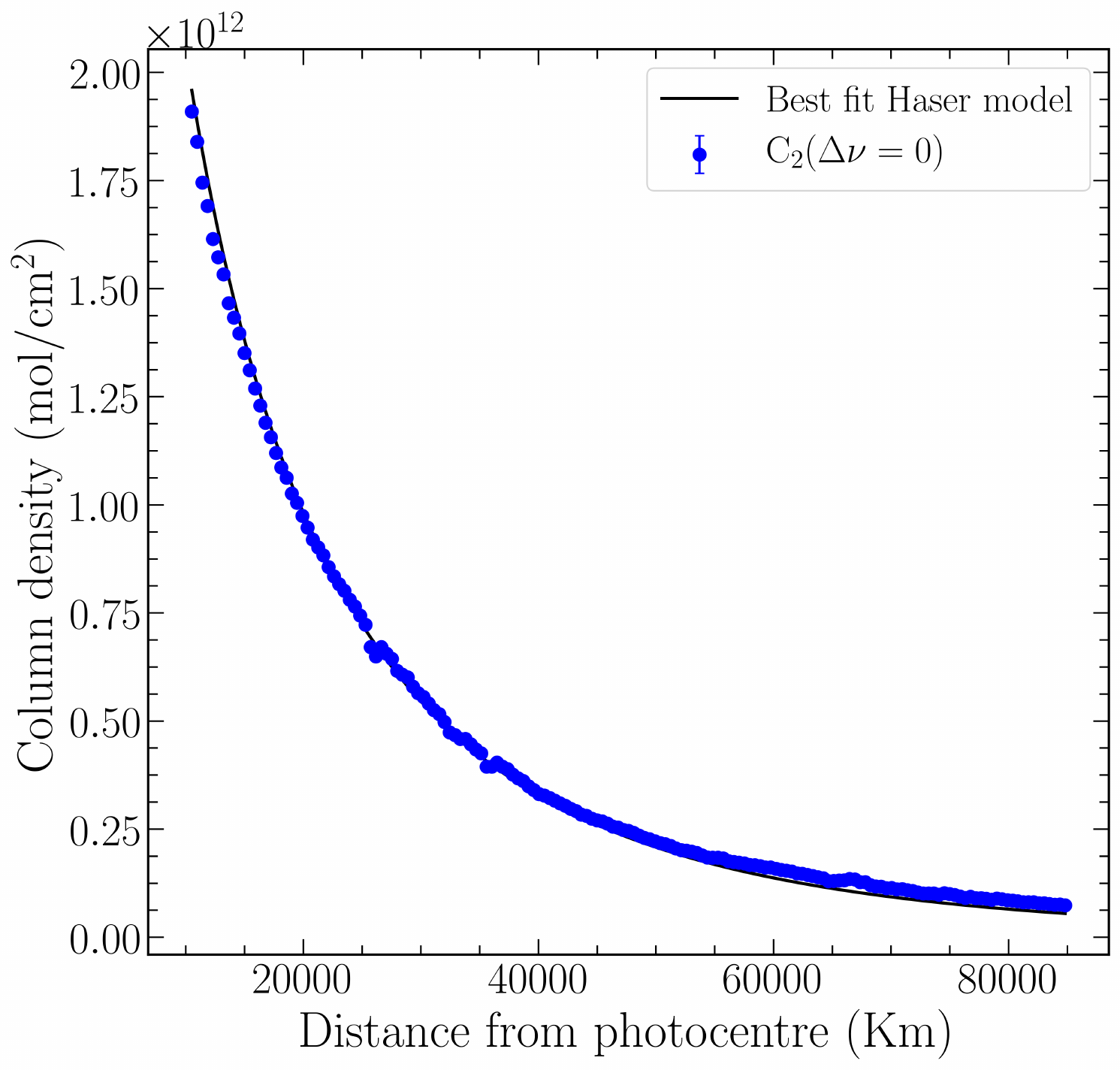}
   \includegraphics[scale=0.2]{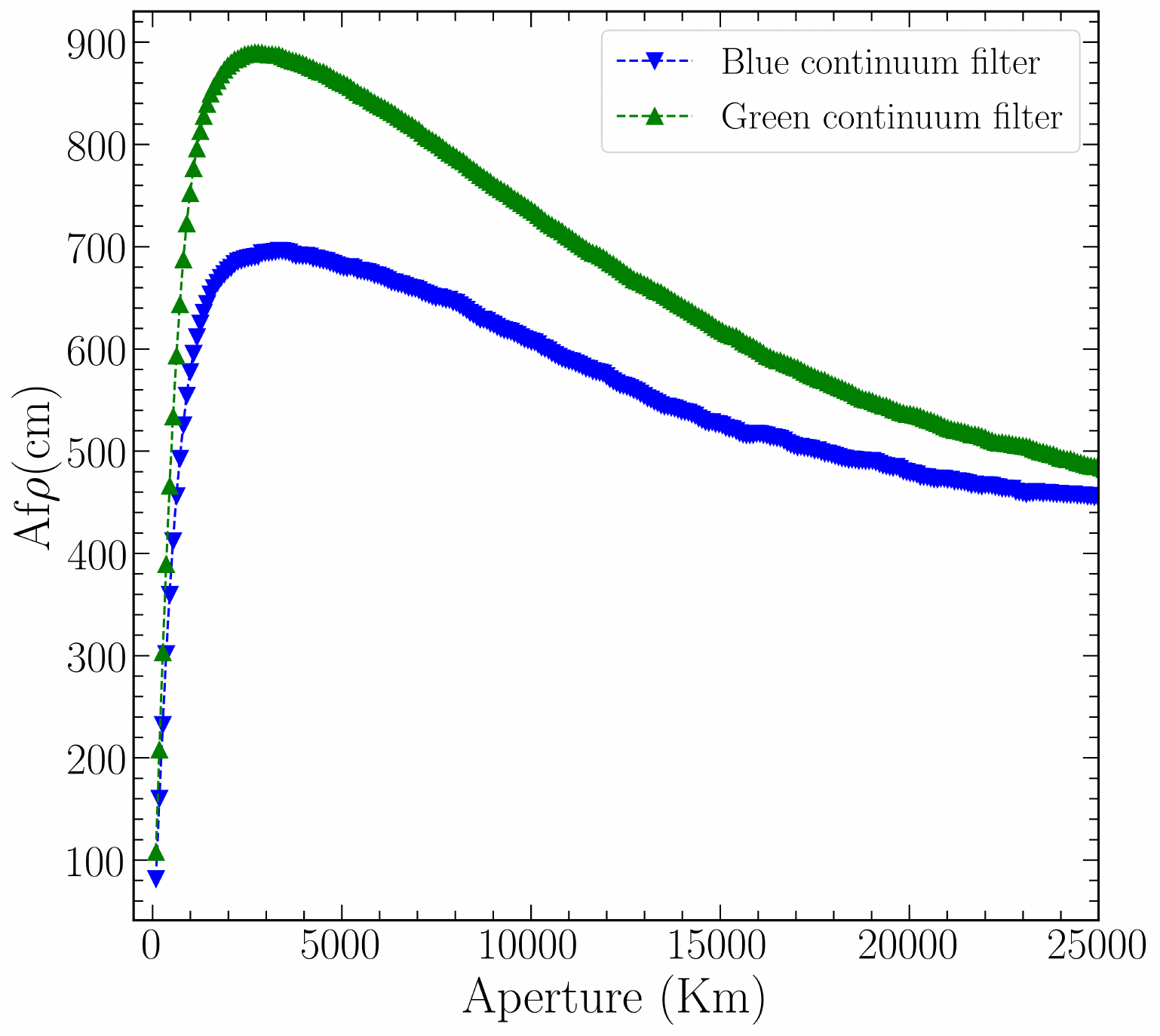}
\begin{minipage}{12cm}
\caption{Depiction of the column density profiles of various molecules observed in comet C/2020 F3 (\textit{Left panel}), column density profile of C$_2$ molecule in comet C/2020 F3 overplotted with the best fit Haser model (\textit{Middle panel}) and the characteristic Af$\rho$ profile of comet 67P in two different continuum bands (\textit{Right panel}).}
\label{longslit}
\end{minipage}
\end{figure}

\begin{figure}[h!]
    \centering
    \includegraphics[width=0.45\linewidth]{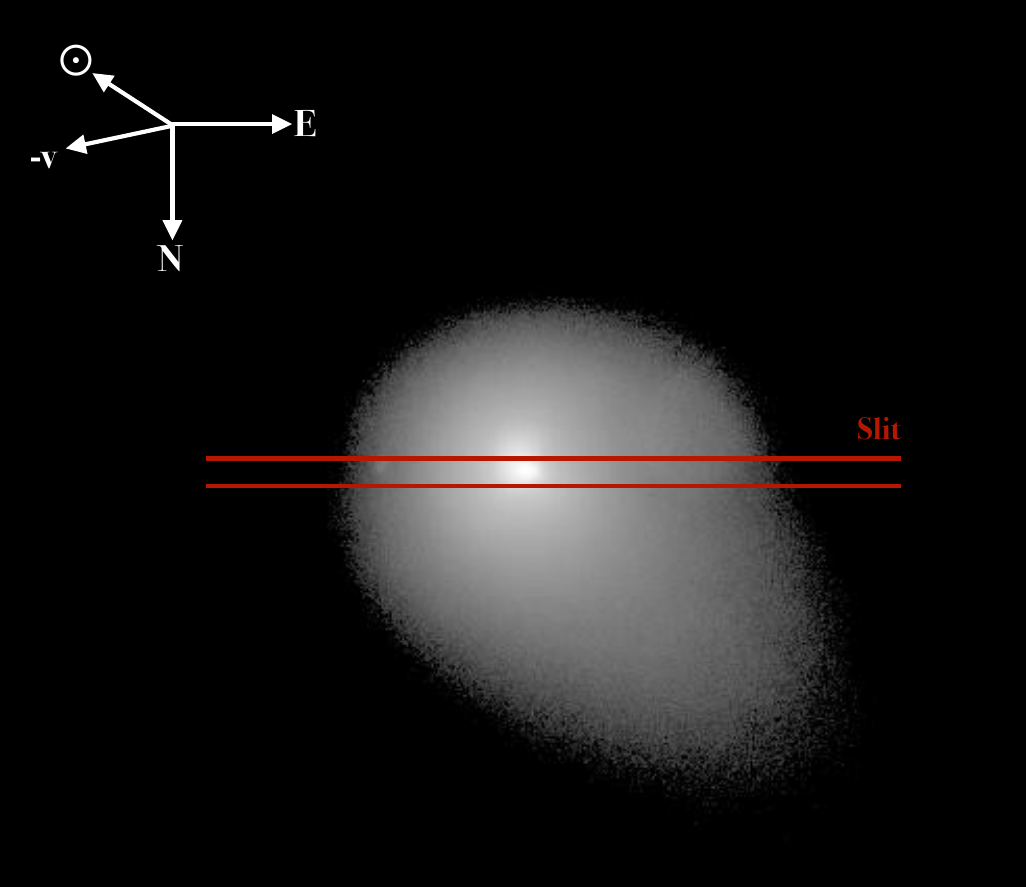}
    \begin{minipage}{12cm}
    \caption{Illustration of the orientation of the long slit in the HFOSC instrument.}
    \label{F3_slit}
    \end{minipage}
\end{figure}

\begin{figure}[h!]
    \centering
    \includegraphics[width=0.75\linewidth]{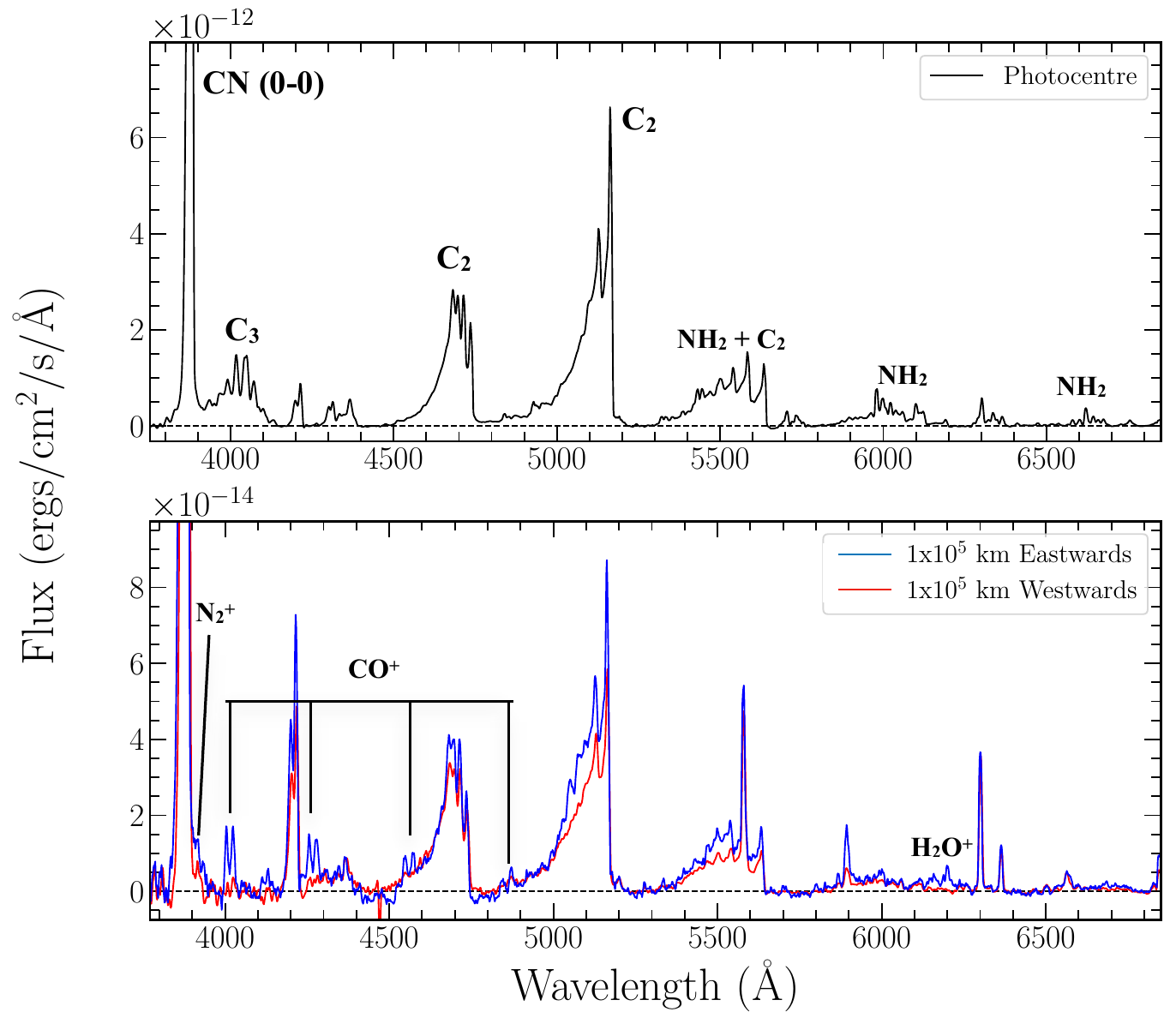}
    \begin{minipage}{12cm}
    \caption{Comparison of spectra extracted from photocentre (\textit{Top panel}) and from the extreme points on the slit (\textit{Bottom panel}) for comet C/2020 F3 observed on 2020-07-24 using the HFOSC instrument on HCT}.
    \label{F3ions}
    \end{minipage}
\end{figure}

\begin{figure}[h!]
    \centering
    \includegraphics[width=0.8\linewidth]{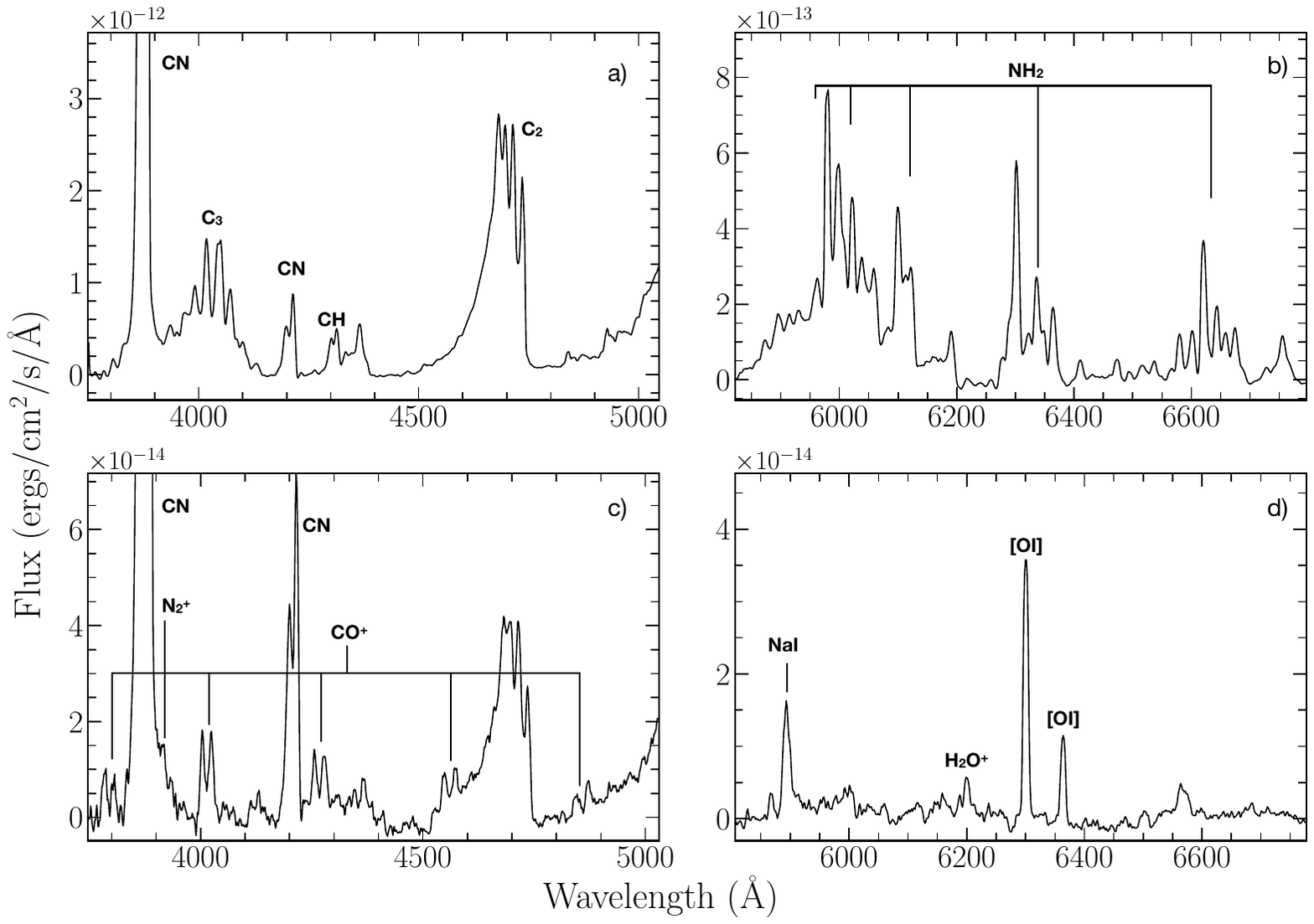}
    \begin{minipage}{12cm}
    \caption{Spectra of comet C/2020 F3,observed on 2020-07-24 using the HFOSC instrument on HCT, at photocentre (\textit{(a) and (b)}) and 10$^5$ km tailward (\textit{(c) and (c)}).}
    \label{F3_centre_east}
    \end{minipage}
\end{figure}

Spectroscopy with a long slit makes this technique more beneficial in the case of comet observations. As comets are extended objects, they have spatial variations in the spectrum, as seen in the left panel of Figure \ref{rawdata}. The spatial variation of each of the molecular emissions seen in a comet spectrum (as shown in Figure \ref{fig:46P}) can be used to extract their column density profile as a function of distance from the photocentre (see left panel of Figure \ref{longslit}). Column density profiles of the major molecules can then be used to extract their production rates (molecules/s) by implementing the Haser model \citep{haser} (see middle panel of Figure \ref{longslit}), as explained in \cite{156P_aravind}. The spatial variation of flux in the continuum bands displayed in the right panel of Figure \ref{comet_imaging} can be used to compute the characteristic profile of the dust proxy parameter Af$\rho$, as shown in the right panel of Figure \ref{longslit}, as explained in \cite{156P_aravind}.\\
Long-slit spectroscopy also provides us with an opportunity to examine the variation of emissions in the cometary spectrum from one end of the slit to the other. This means at each location of the slit, we would be probing emissions from different parts of the coma or even the tail. In earlier days, \cite{Halley_bradfield_ions} observed comets Halley (1986 III) and Bradfield (1987 XXIX) both at the photocentre and in the tailward direction to find differences in the emission. They were able to spot emissions from CO$^+$ and N$_2^+$ only in the tailward spectrum as they would be suppressed by the major emission at the photocentre. Later, \cite{halley_spatial_ions} also did similar work in a more sophisticated manner to observe comet Halley at different spatial locations in the coma to detect ionic emissions. Such observational techniques are important in detecting the ionic species, scarcely detected in cometary spectra, to better understand the composition of these icy bodies. But it is not so straight forward to place the slit at certain locations of the coma to get hold of the ionic emissions.\\
In certain cases, if the ion tail of the comet coincides with the orientation of the slit, we will be able to detect ionic emissions at the extreme parts of the slit (if the slit is long enough), less dominated by the major emissions. In the case of the Hanle Faint Object Spectrograph Camera (HFOSC) instrument on HCT or the ARIES-Devasthal Faint Object Spectrograph $\&$ Camera (ADFOSC) instrument on DOT, the slits are long enough, about 11 arcmin and 8 arcmin respectively, to probe very large distances of the coma from the photocentre. Figure \ref{F3_slit} depicts the spatial direction of the comet C/2020 F3 probed through the long slit available in the HFOSC instrument.  \\
During comet observations, the slit on these instruments can also be rotated at any angle so as to coincide its spatial axis with the tail to probe emissions along the tail of the comet. The default orientation of the slit was used for the observation of comet C/2020 F3 in July 2020. The spectrum extracted for the photocentre of the comet seemed to be similar to that expected for any Solar system comet with regular emissions from CN, C$_3$, C$_2$, NH$_2$ etc. (see top panel in Figure \ref{F3ions}). The long slit in the instrument facilitated the extraction of the spectrum at distances of $\sim$10$^5$ km to both sides from the photocentre. Surprisingly, the spectrum extracted at the extreme point of the slit, Eastwards from the photocentre, showed emissions from ionic species like CO$^+$, N$_2^+$ and H$_2$O$^+$. At the same time, the spectrum extracted at a similar distance Westwards of the photocentre lacked these emissions confirming their cometary origin. Figure \ref{F3ions} compares the spectrum extracted from the photocentre as well as from the extreme points on the slit, clearly depicting the difference in emissions.
The column density profile of comet C/2020 F3 shown in the left panel of Figure \ref{longslit} clearly demonstrates that the C$_3$ and NH$_2$ emissions die down significantly at distances of about 20,000 km from the photocentre. The drastic difference in the emissions observed in the spectra extracted from the photocentre and the extreme Eastward point on the slit is clearly illustrated in Figure \ref{F3_centre_east}. While comparing the spectra extracted from the extreme points on the slit, Westwards and Eastwardss, it was observed that the intensity of the Oxygen lines (6300 \AA~and~6364 \AA) matched very well, confirming its origin to be Telluric. At the same time, the NaI doublet lines observed at 5890/5896 \AA~(seen as a single line due to the low resolution) showed a difference in intensities, wherein the Eastward spectra showed higher intensity (see Figure \ref{F3ions}). \\
During the follow-up spectroscopic observations of the comet by \cite{F3_followup}, the authors illustrate that the NaI doublet emissions had gone extremely faint after 23$^{rd}$ July 2020. But, the difference in the intensities between the emissions observed at the two extreme points on the slit raises a possibility for the emissions at the Eastward end to be of cometary origin. The origin of this sodium emission in bright comets, when they are within a heliocentric distance of about 0.8 AU, has been linked to the dust particles present in the coma and dust tails \citep{swamy}. Hence, the presence of NaI emissions in our observations, which pops out as the major NH$_2$ emissions subside, could be due to the dust particles in the extended coma. 

\section{Summary}
In this article, it has been briefly put forward that spectroscopy is an efficient tool to have a simultaneous study of the gas and emissions arising across the optical range from a comet, thereby reducing the possibility of effects from temporal changes in the emissions. In order to have the complete access to all the emissions, 380-700 nm is an optimal requirement for cometary studies in low-resolution spectroscopy. The availability of long slit aids in analysing the spatial profile of various molecule's column density and hence makes the computation of production rates and parent/daughter scale length modelling possible. In certain cases, the critical analysis of long split spectra of comets could provide immense information regarding the spatial variation of these emissions. During observations, a live slit viewer with the possibility of manual position correction or precise sidereal tracking methods is essential to obtain long exposures, thereby providing better SNR in the spectra. With more compositional studies on comets, we would get to know these interesting minor bodies better and decipher the processes that occurred during the initial phases of the evolution of our Solar system.

\begin{acknowledgments}
We acknowledge the local staff at the Mount Abu Observatory, the DOT observatory and the staff of the Indian Astronomical Observatory, Hanle and Centre For Research \& Education in Science \& Technology, Hoskote that made these observations possible. The facilities at IAO and CREST are operated by the Indian Institute of Astrophysics, Bangalore. The facilities at DOT are operated by ARIES, Nainital. \\Work at Physical Research Laboratory is supported by the Department of Space, Govt. of India.
\end{acknowledgments}

\begin{furtherinformation}

\begin{orcids}
\orcid{0000-0002-8328-5667}{K}{Aravind}
\orcid{0000-0002-7721-3827}{Shashikiran}{Ganesh}
\end{orcids}

\begin{authorcontributions}
This work is a result of equal contribution from all the authors.
\end{authorcontributions}

\begin{conflictsofinterest}
The authors declare no conflict of interest.
\end{conflictsofinterest}

\end{furtherinformation}

\bibliographystyle{bullsrsl-en}

\bibliography{reference}

\end{document}